\documentclass[sn-mathphys-num]{sn-jnl}% Math and Physical Sciences 

\usepackage{graphicx}%
\usepackage{multirow}%
\usepackage{amsmath,amssymb,amsfonts}%
\usepackage{amsthm}%
\usepackage{mathrsfs}%
\usepackage[title]{appendix}%
\usepackage{textcomp}%
\usepackage{manyfoot}%
\usepackage{booktabs}%
\usepackage{algorithm}%
\usepackage{algorithmicx}%
\usepackage{algpseudocode}%
\usepackage{listings}%

\theoremstyle{thmstyleone}

\theoremstyle{thmstyletwo}

\theoremstyle{thmstylethree}

\begin{document}

\title[Article Title]{Effect of Parametric Variation of Chordae Tendineae Structure on Simulated Atrioventricular Valve Closure}

\author[1]{\fnm{Nicolas R.} \sur{Mangine}}\email{manginen@chop.edu}
\equalcont{These authors contributed equally to this work.}
\author[1,2]{\fnm{Devin W.} \sur{Laurence}}\email{laurenced@chop.edu}
\equalcont{These authors contributed equally to this work.}
\author[1]{\fnm{Patricia M.} \sur{Sabin}}\email{sabinp@chop.edu}

\author[1,2]{\fnm{Wensi} \sur{Wu}}\email{wuw4@chop.edu}

\author[1]{\fnm{Christian} \sur{Herz}}\email{herzc@chop.edu}

\author[1]{\fnm{Christopher N.} \sur{Zelonis}}\email{zelonisc@chop.edu}

\author[1,2]{\fnm{Justin S.} \sur{Unger}}\email{ungerj@chop.edu}

\author[3]{\fnm{Csaba} \sur{Pinter}}\email{csaba.pinter@ebatinca.com}

\author[4]{\fnm{Andras} \sur{Lasso}}\email{lasso@queensu.ca}

\author[5]{\fnm{Steve A.} \sur{Maas}}\email{steve.maas@utah.edu}

\author[5]{\fnm{Jeffrey A.} \sur{Weiss}}\email{jeff.weiss@utah.edu}

\author*[1,2]{\fnm{Matthew A.} \sur{Jolley}}\email{jolleym@chop.edu}

\affil[1]{\orgdiv{Jolley Lab, Department of Anesthesia and Critical Care Medicine}, \orgname{Children's Hospital of Philadelphia}, \city{Philadelphia}, \state{PA}, \country{US}}

\affil[2]{\orgdiv{Division of Cardiology}, \orgname{Children's Hospital of Philadelphia}, \city{Philadelphia}, \state{PA}, \country{USA}}

\affil[3]{\orgname{EBATINCA}, \city{Las Palmas de Gran Canaria}, \state{Las Palmas}, \country{Spain}}

\affil[4]{\orgname{Queens University}, \state{Ontario}, \country{CA}}

\affil[5]{\orgname{University of Utah}, \city{Salt Lake City}, \state{Utah}, \country{USA}}

\abstract{\textbf{Purpose:} Many approaches have been used to model chordae tendineae geometries in finite element simulations of atrioventricular heart valves. Unfortunately, current ``functional" chordae tendineae geometries lack fidelity (e.g., branching) that would be helpful when informing clinical decisions. The objectives of this work are (i) to improve synthetic chordae tendineae geometry fidelity to consider branching and (ii) to define how the chordae tendineae geometry affects finite element simulations of valve closure. 

\textbf{Methods:} In this work, we develop an open-source method to construct synthetic chordae tendineae geometries in the SlicerHeart Extension of 3D Slicer. The generated geometries are then used in FEBio finite element simulations of atrioventricular valve function to evaluate how variations in chordae tendineae geometry influence valve behavior. Effects are evaluated using functional and mechanical metrics.

\textbf{Results:} Our findings demonstrated that altering the chordae tendineae geometry of a stereotypical mitral valve led to changes in clinically relevant valve metrics (regurgitant orifice area, contact area, and billowing volume) and valve mechanics (first principal strains). Specifically, cross sectional area had the most influence over valve closure metrics, followed by chordae tendineae density, length, radius and branches. We then used this information to showcase the flexibility of our new workflow by altering the chordae tendineae geometry of two additional geometries (mitral valve with annular dilation and tricuspid valve) to improve finite element predictions.

\textbf{Conclusion:} This study presents a flexible, open-source method for generating synthetic chordae tendineae with realistic branching structures. Further, we establish relationships between the chordae tendineae geometry and valve functional/mechanical metrics. This research contribution helps enrich our open-source workflow and brings the finite element simulations closer to use in a patient-specific clinical setting.
}

\maketitle

\section{Introduction}\label{sec:introduction}

Atrioventricular valve (AVV) disease is associated with significant morbidity and mortality in both the developing and developed world. Mitral valve (MVR) and tricuspid valve regurgitation (TVR) are thought to affect $5$\,million~\cite{kolte2020current} and $1.6$\,million~\cite{henning2022tricuspid} people in the U.S., respectively. A healthy MV and TV will adequately control the flow of blood between the atria and ventricles on the left and right sides of the heart, respectively. Consequently, failure of these AVVs can result in inefficient blood circulation leading to end-organ dysfunction (e.g., liver, lung, kidney), or death. MVR is associated with an increased risk of congestive heart failure and stroke and is the most prevalent form of AVV failure~\cite{howsmon2020mitral, sannino2017survival}. The TV, while previously described as the ``forgotten valve", has increasingly been the subject of important investigations~\cite{hahn2022tricuspid} as part of an effort to better treat TV dysfunction.  AVV failure is similarly devastating to both children and adults with congenital heart disease, and in particular to single ventricle patients~\cite{ho2020left, king2019atrioventricular}.

\begin{figure}[h]
\includegraphics[width=0.95\textwidth]{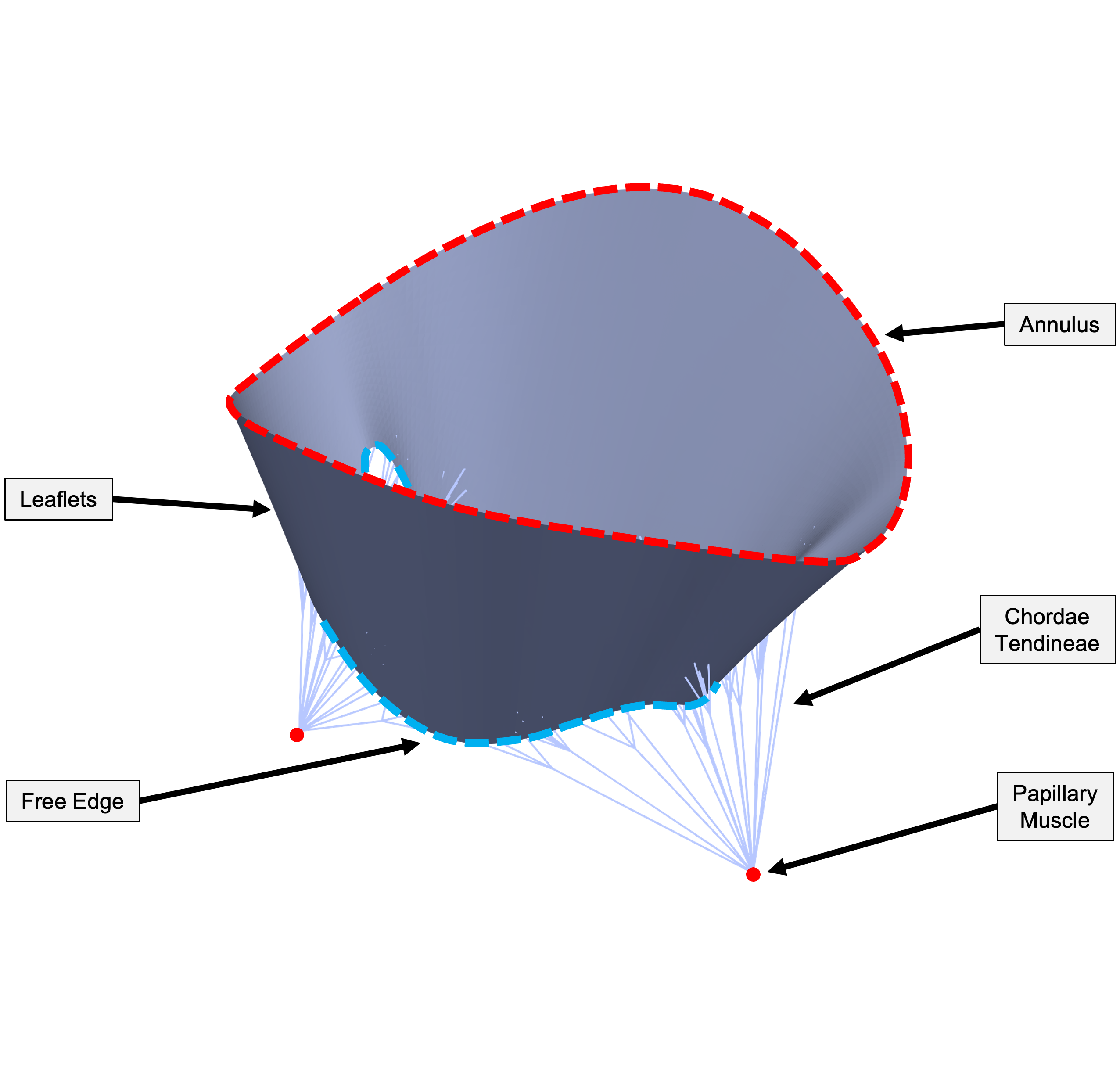}
\centering
\caption{Unloaded mitral valve with key anatomical features.}
\label{Figure: Mitral Anatomy}
\end{figure}

The core anatomy of an AVV consists of leaflets, a ring-shaped annulus attaching the leaflets to the heart wall, and chordae tendineae (CDT) that anchor the leaflets to the papillary muscles (PM) in the ventricle wall (Fig.~\ref{Figure: Mitral Anatomy}). Morphological changes or failure of any of these components can lead to an incompetent valve that allows blood to flow backward from the ventricle to the atrium during ventricular contraction, leading to atrioventricular valve regurgitation (AVVR)~\cite{wu2018incidence, mozaffarian2016heart}. Based on etiology, the factors of AVVR have been categorized as either \textit{primary} (e.g., direct) or \textit{secondary} (e.g., indirect) in terms of its effects on the AVV~\cite{sannino2017survival, howsmon2020mitral, iddawela2022paediatric}. Myxomatous valve degeneration is an example of primary AVVR that is characterized by elongated CDT and prolapsing leaflets stemming from decreased stiffness of both CDT and leaflet tissue~\cite{gupta2009abundance, barber2001mechanical}. In contrast, ischemic cardiomyopathy is an example of secondary AVVR related to left ventricular dilation and remodeling~\cite{howsmon2020mitral,beeri2008mitral,liel2000design}.

In both primary and secondary AVVR cases, impaired CDT can contribute to valve dysfunction and failure given the essential role they play in normal AVV function~\cite{nam2022visualization, montanhesi2022simplifying}. In primary AVVR, CDT rupture can result in leaflet prolapse. While in secondary cases, elevated chordal tension in the setting of ventricular dilation, or ischemia-related alterations to the ventricle, can lead to CDT restriction of leaflet motion and reduction in leaflet contact (coaptation). Congenitally abnormal CDT structures in children, such as those seen in parachute or arcade AVV~\cite{khurana2024tale, fritz2020mitral}, play a deleterious role in valve function.  Consequently, chordal tension and direction plays a significant part in the successful repair of complex congenital heart disease with valvular abnormalities such as complete atrioventricular canal and hypoplastic left heart syndrome~\cite{nam2022dynamic, nam2022visualization, ho2020left, bautista2014, spinner2011vitro, yamauchi2012right}.

Given the essential role of CDT in valve function, significant effort has been made to develop therapies directed at replacing or altering chordal geometry to achieve improved AVV function. These interventions are dependent on CDT mechanics, especially as related to their physical classification. Specifically, CDT classified as ``primary" are inserted within leaflet free edge and prevent leaflet prolapse during ventricular contraction. Secondary CDT, embedded in leaflet body, support the mechanical interaction between AVV and ventricle, with additional stability within the leaflet basal area provided by tertiary CDT~\cite{salik2023mitral, ross2020mechanics, rodriguez2004importance, obadia1997mitral}. Thus, precise assessment of insertion location, and optimal chordal length, is necessary to understand valve function and to achieve desired therapeutic outcomes. This includes surgical interventions such as: neochordae reconstruction where artificial primary CDT are created to restore function to a prolapsing valve~\cite{montanhesi2022simplifying}; and PM relocation where CDT-to-leaflet insertion vectors are modified to improve valve function~\cite{kong2018repair, spinner2011vitro, yamauchi2012right, dreyfus2006papillary, rabbah2013effects}).

In this context, realistic representation and fidelity of CDT topology is also critical to emerging computational simulations of AVV function~\cite{sacks2019simulation}. Notably, finite element (FE) and other multi-physics simulations of AVV function are uniquely positioned to investigate mechanisms of valve dysfunction.  In addition, when combined with patient-specific three-dimensional (3D) images (e.g., 3D echocardiography, computed tomography, cardiac magnetic resonance imaging), these computational techniques have the potential to inform iterative preoperative optimization of valve repair ``in silico"~\cite{sacks2019simulation, haese2024impact, wu2023effects, laurence2020pilot, mathur2022texas}. However, while advances in 3D imaging provide a reliable and non-invasive means of determining valve leaflet structure~\cite{qureshi2019tricuspid, ryan2006emerging, muraru20193, nam2023modeling, nam2022visualization, lasso2022slicerheart}, modern clinical imaging cannot yet reliably resolve CDT geometry at their individual chordal insertions with the accuracy and 
fidelity of ``ground truth" geometries achieved from \textit{ex vivo} and pathological specimens~\cite{sacks2019simulation, khalighi2019development}.

As a result, significant effort has gone into creating image-informed~\cite{kong2018finite, pham2017finite, kong2018repair, kong2020finite} or functionally equivalent~\cite{khalighi2019development} CDT topology for the computational simulation of valve function. Most relevant to the present work, Khalighi \textit{et al}.~\cite{khalighi2019development} demonstrated that CDT geometries could be simplified to random insertions of CDT-like structures without branching. These functionally equivalent CDT topologies provided similar predictions as CDT geometries created from high-fidelity \textit{ex vivo} images of ovine hearts. Although the simplified CDT geometries could predict valve behavior, they do not have realistic features (i.e., branching) necessary to model CDT-related pathology or interventions (e.g., chordal rupture or neochordae). As such, others have created FE models with various implementations of branching CDT geometries~\cite{mathur2022texas, laurence2020pilot, johnson2021parameterization}. However, the sensitivity of FE simulations of valve function to variations in chordal branching patterns and distributions remains unknown.  Furthermore, there is not an existing procedural method for the precise parametric and reproducible creation of specific chordal branching patterns, geometries and density. 

The objectives of this work are thus to (i) define how CDT geometry affects FE predictions of AVV closure and (ii) to compare FE simulation results of the more realistic branched CDT geometries to simpler non-branching CDT geometries previously used by ourselves~\cite{wu2022computational, wu2023effects} and others~\cite{khalighi2019development}. In order to achieve this analysis we  created a novel open-source module in SlicerHeart to procedurally generate chordal branching geometries~\cite{lasso2022slicerheart}.  We then employed this framework to quantify how variation of five CDT parameters influenced clinically relevant metrics of valve closure metrics in three realistic atrioventricular valve models.

\section{Methods}\label{sec:methods}

\subsection{Geometry Creation}\label{sec:geometry}

\begin{figure}[h]
\includegraphics[width=0.95\textwidth]{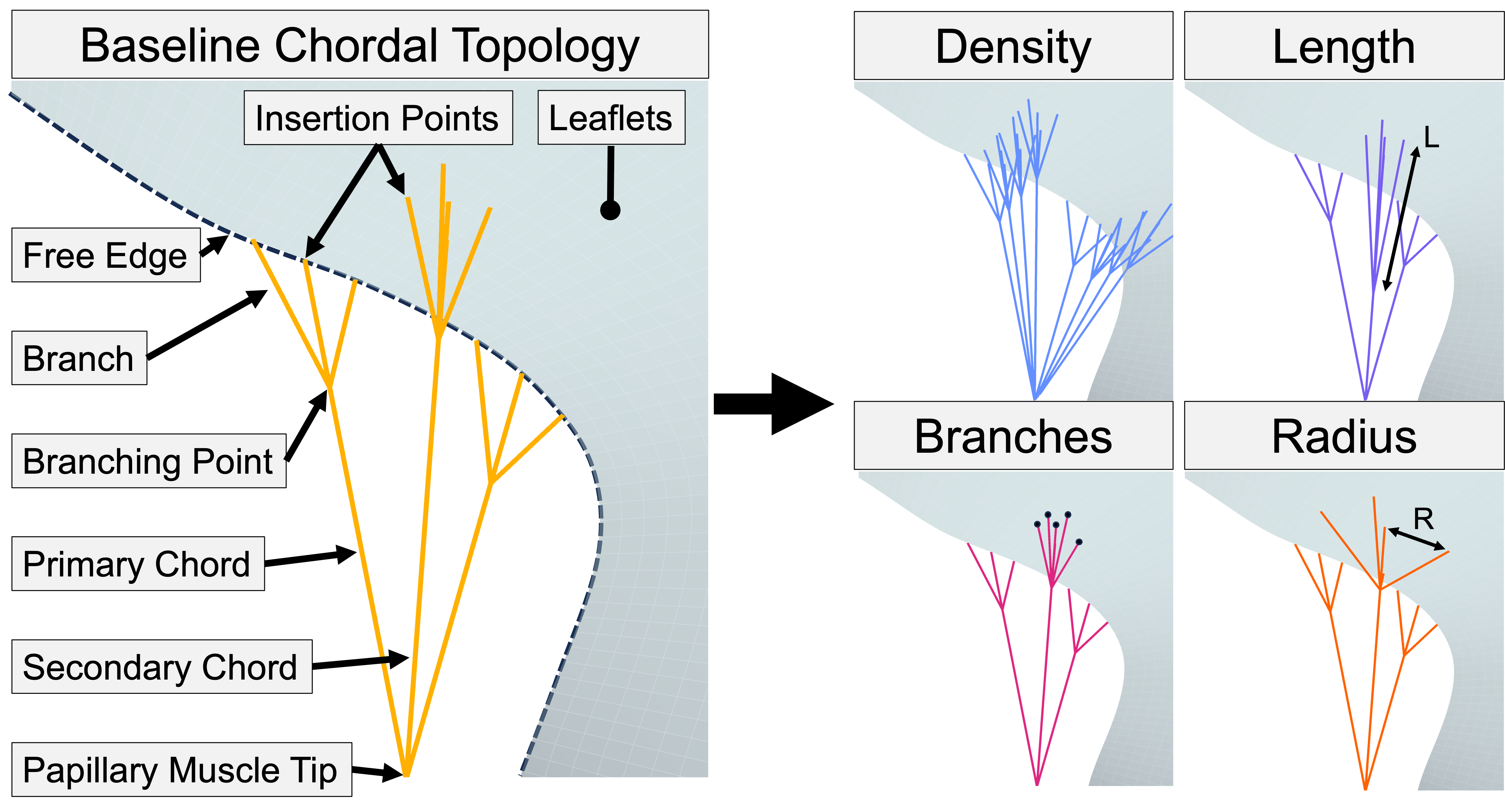}
\centering
\caption{Method for varying the parameters of the chordae tendineae. Baseline topology can be modified by CDT density, length, number of branches, and branching radius.}
\label{Figure: Parametric Tools}
\end{figure}

Three synthetic AVV geometries were created to understand how CDT geometry influences valve closure: a stereotypical MV geometry, the stereotypical MV geometry with a dilated annulus, and a stereotypical TV. The Non-Uniform Rational B-Spline (NURBS) surfaces representing the leaflets were used in our new parametric CDT creation tool to procedurally generate CDT geometries with branching structures. Specific details regarding the study scenarios are provided in Section~\ref{sec:StudyScenarios}. This tool was implemented as a scripted Python module of 3D Slicer within the SlicerHeart extension. The module offers a comprehensive user interface and the necessary logic functions for CDT generation based on user selection. Step-wise schematics are provided in Figure~\ref{Figure: Parametric Tools}. First, a finely meshed leaflet surface is generated for CDT geometry creation. Next, the PM tips, valve surface, and CDT insertion regions are selected in the module interface. The user then defines CDT geometry parameters, including: CDT type (primary, secondary, or both), insertion density, number of branches, branching length, and branching radius. Finally, the CDT are generated with chords emanating from the closest PM tip into each insertion region (See Video S1 in the Supplementary Material section for the procedure). This CDT creation tool is implemented in the SlicerHeart extension (www.slicerheart.org)~\cite{lasso2022slicerheart} for 3D Slicer~\cite{fedorov2012} as the \texttt{ValveFEMExport} module. All code is available open source at www.github.com/SlicerHeart.

\subsection{Finite Element Simulation}\label{sec:FEMethods}

FE analysis of the valve closure was performed using our established workflow in FEBio~\cite{wu2022computational, wu2023effects, maas2012febio}. The MV and TV CDT were modeled as newly implemented linear truss elements with a prescribed modulus of $150$\,MPa and cross sectional area (CSA) of $1.0$, and $0.44$\,${\rm mm}^2$~respectively in FEBio~\cite{ross2020mechanics}. The leaflets were discretized using 4-node linear quadrilateral shell elements (\texttt{quad4}) with assumed thickness of $0.396$\,{mm}~\cite{wu2023effects}. The nonlinear material behavior of the leaflets was modeled via the nearly incompressible isotropic Lee-Sacks constitutive model~\cite{lee2014} with strain energy density ($\psi$) defined as:

\begin{equation}
\label{Eq:LeeSacks}
    \psi = \frac{c_0}{2}\left(I_{1}-3\right) + \frac{c_1}{2}\left(e^{c_{2}\left(I_{1}-3\right)^2}-1\right)
\end{equation}

\noindent Here, $c_0=200$\,kPa, $c_1=2968.4$\,kPa, $c_2=0.2661$ are material coefficients for the MV~\cite{kamensky2018contact} and $c_0=10$\,kPa, $c_1=0.209$\,kPa, $c_2=9.046$ are material coefficients for the TV. Near incompressibility was enforced with a bulk modulus of $5000$\,kPa and $10000$\,kPa for the MV and TV, respectively. $I_1$ is the first invariant of the right Cauchy deformation tensor $\mathbf{C}=\mathbf{F}^T\mathbf{F}$, and $\mathbf{F}$ is the deformation gradient. Zero displacement boundary conditions were applied to the annulus and PM tip nodes. A linearly increasing systolic pressure ($100$ and $23.7$\,mmHg for the MV and TV respectively) was applied to the ventricular leaflet surface over $0.005$\,s. Leaflet contact was modeled using a potential-based formulation~\cite{kamensky2018contact}. Dynamic analysis was performed using the generalized-alpha integration rule, mass damping for both the leaflets and CDT ($C=5000$\,Ns/m), and the FEBio automatic time stepping algorithm ($\Delta t_{\rm min} = 10^{-8}$\,s and $\Delta t_{\rm max} = 10^{-4}$\,s). Quasi-static analysis was sought but dynamic analyses were performed to achieve better simulation convergence~\cite{wu2022computational}. 

\subsection{Study Scenarios}\label{sec:StudyScenarios}

We performed several investigations to understand how CDT geometry influences FE predictions of AVV function. First, we used our existing synthetic MV geometry~\cite{laurence2024} to compare the refined CDT geometry generated using our new approach to the simplified CDT geometry used in our prior work and that of others. The ``baseline" MV CDT geometry was then parametrically varied to understand how changes to the CDT structure influence AVV closure. Finally, the understanding gained via the synthetic MV geometry was extended to generate CDT geometries for two additional AVVs: one MV with a dilated annulus and one TV. 

\subsubsection{Establishing Baseline CDT Geometry for the Synthetic MV}\label{Baseline CDT}
A stereotypical, healthy MV geometry was created to evaluate the effect of CDT geometry on valve closure (see Section~\ref{sec:geometry}). Our prior CDT generation workflow~\cite{wu2022computational} (based on the work of Khalighi \textit{et al}.~\cite{khalighi2019development}) generated a simplified CDT geometry with $115$ insertions. We successfully used our new parametric workflow (see Section~\ref{sec:geometry}) to obtain $114$ insertion points with the following parameters: primary CDT density of $2$\,${\rm chords}/{\rm cm}$, secondary CDT density of $3$\,${\rm chords/cm}^2$, branch length of $3.5$\,mm, number of branches as $3$ and $4$ for primary and secondary CDT respectively, and insertion radius of $1$\,mm. We compared the FE simulation results using these two CDT geometries (fully simplified and branched) as a first step towards understanding how increased CDT fidelity may influence the FE predictions. It should also be noted that CDT density is quantified in our study in terms of leaflet free edge insertion distance (${\rm chords}/{\rm cm}$) for primary chords, and leaflet surface area of insertion (${\rm chords/cm}^2$) for secondary chords.

\subsubsection{Mesh Convergence Study}\label{Mesh Density}
We performed an initial study to understand how mesh density may influence our findings. In brief, the FE mesh for the synthetic MV was refined using HyperMesh (Altair, Troy, MI), resulting in a wide range of FE mesh densities ($2500$-$20000$ elements). The MV function was simulated using our FE simulation workflow in FEBio (Section~\ref{sec:FEMethods}). The functional and mechanical metrics (see Section~\ref{sec:Metrics}) were compared across mesh densities. 

\subsubsection{Parametric Variations of Primary CDT Geometry}\label{Primary CT}
The baseline MV was then used to evaluate primary CDT influence on valve closure. The primary CDT density ($1$-$3$\,${\rm chords}/{\rm cm}$), branch length ($1$-$10$\,mm), branch quantity ($2$-$5$), and branching radius ($0.4$-$1.8$\,mm) values were varied using the \texttt{ValveFEMExport} module (Section~\ref{sec:geometry}). These ranges were chosen based on the minimum available values in the SlicerHeart module and the point when chordae insertion points began overlapping (except for branch length). Additionally, we varied the primary CDT CSA ($0.2$-$2.0$\,${\rm mm}^2$) in FEBio based on values reported by Ross \textit{et al}.~\cite{ross2020mechanics}. 

\subsubsection{Parametric Variations of Secondary CDT Geometry}\label{Secondary CT}

After establishing the role of primary CDT, we used the baseline MV to evaluate how secondary CDT influence valve closure. Both primary and secondary CDT were included based on preliminary simulations that showed primary CDT were needed to prevent localized leaflet free edge prolapse. We varied the secondary CDT density ($1$-$10$\,${\rm chords/cm}^2$), branch length ($1$-$10$\,mm), branch quantity ($1$-$5$), and branch radius ($0.4$-$3$\,mm). Similar to the primary CDT study, the minimum and maximum values were selected based on minimum values available in the SlicerHeart module and when CDT insertions began overlapping, respectively. Finally, the secondary CDT CSA was varied in FEBio from $0.2$-$2.0$\,${\rm mm}^2$, while keeping the primary CDT constant at $1.0$\,${\rm mm}^2$ based on values reported by Ross \textit{et al}.~\cite{ross2020mechanics}.

\subsubsection{Extension to Additional AVV Geometries}
Two additional valves were created using the procedure outlined in Section~\ref{sec:geometry}: a MV with annular dilation (ADMV) and a TV. The ADMV was created by uniformly increasing the annulus of the healthy, synthetic MV by $10$\% while not altering the leaflet and commissure heights. The TV was adapted from Wu \textit{et al}.~\cite{wu2023effects} and a generalized annulus was created from~\cite{besler2018transcatheter}. Neither valve had a reference for CDT topology so the baseline CDT values from the normal MV were used and valve closure metrics were evaluated. Using the trends observed in the normal MV CDT studies, the CDT parameters were then updated to create more realistic valve closure. This qualified as localized free edge leaflet prolapse for the ADMV and unrealistic regurgitant orifice area (ROA) for the TV. 

\subsection{Analysis}\label{subsec5}

\begin{figure}[h]
\includegraphics[width=0.80\textwidth]{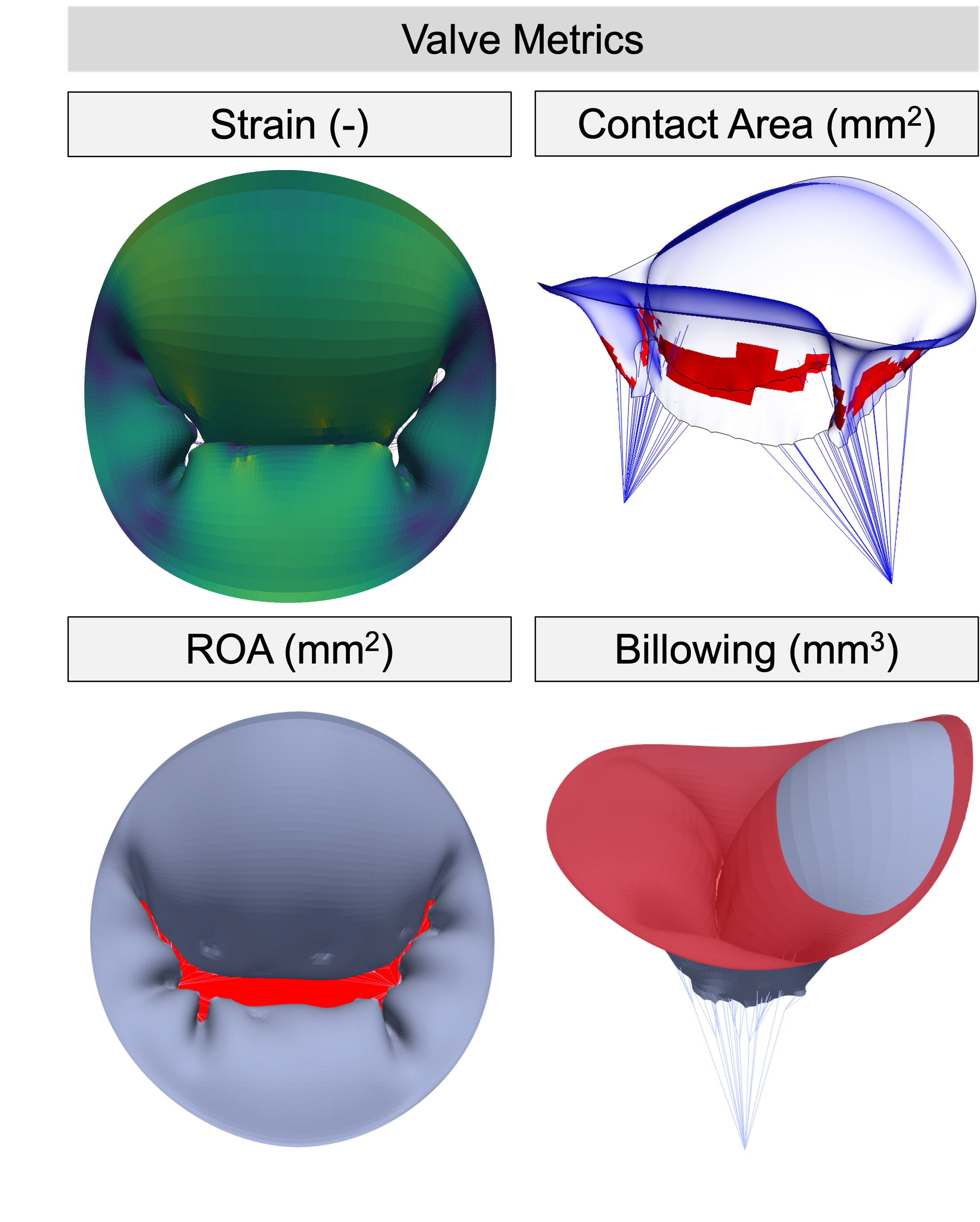}
\centering
\caption{Four valve metrics were identified to quantify valve closure. The medial surface of the mitral valve in its closed state is used to visualize strain, contact area (CA), and billowing volume, while a dilated geometry is used to show the regurgitant orifice area (ROA). Strain measurements are represented with a color map, CA, and ROA are shown in red, and the billowing volume is shown as the leaflet above the red annulus plane.}
\label{Figure: Metrics}
\end{figure}

\subsubsection{Mechanical and Functional Metrics}\label{sec:Metrics}

We evaluated the effect of CDT geometry using both functional and mechanical metrics. The ROA, contact area (CA), and billowing volume are clinically relevant \textit{functional metrics} related to valve function that can inform valve repair. We used a fully-automated method to accurately quantify the ROA by coupling a shrink-wrapping method with raycasting~\cite{wu2023effects}. The CA was calculated by the surface area of the elements with non-zero traction applied by the contact algorithm~\cite{kamensky2018contact}. Billowing volume was calculated in Slicer's \texttt{Parametric Atrioventricular Valve} module as the volume of leaflet above the annulus plane during ventricular contraction (Fig.~\ref{Figure: Metrics}). Finally, we considered the first principal strain of the leaflets as the \textit{mechanical metric} owing to its relationship with valve remodeling~\cite{howsmon2020mitral}. Leaflet strain visualizations include the median, $25\%$ and $75\%$ percentiles, and minimum/maximum values, whereas functional metric visualizations report the singular quantified value.

The ordinary least squares functionality of the Python \texttt{statsmodels} module was used to determine the presence of linear, quadratic, or exponential relationships between the valve metrics and CDT parameters. We considered three criteria to evaluate regression quality: (i) the coefficient of determination ($R^2$), (ii) the \textit{p}-value of each regression coefficient, and (iii) visual inspection, particularly when considering the quadratic relationship (i.e., is the inflection point used to match the data). These findings were used as a quantitative foundation to interpret our results. Note that the regression coefficients are only relevant to the stereotypical MV geometry, and the trends may not hold when extrapolated due to the nonlinear behavior of the FE simulations. 

\subsubsection{CDT Geometry Comparison}

We compared FE predictions with varying CDT geometries in three ways to understand: (i) how parameters in the SlicerHeart module altered valve closure, (ii) whether CDT insertion density was the central driver of CDT function, and (iii) whether total CDT CSA was the central driver of CDT function. First, we normalized the valve functional metrics to values associated with the baseline CDT geometry---this was repeated for both the primary and secondary CDT geometry studies. Second, we compared functional metrics when varying CDT density and branch number to understand if trends were consistent across different CDT geometries with similar insertion point densities. This analysis aimed to test the hypothesis that CDT insertion density will influence valve behavior regardless of which CDT parameter altered the CDT density. This hypothesis is based on prior work by Khalighi \textit{et al}.~\cite{khalighi2019development} that showed drastically simplified CDT geometries provided functionally equivalent behavior as ground truth CDT geometries from \textit{ex vivo} imaging so long as CDT insertion point density was above a minimum threshold. Finally, we also compared the functional metrics across different CDT geometries with similar total CDT CSA. This analysis sought to test the hypothesis that total CDT CSA will influence valve functional metrics similarly regardless of how the total CDT CSA is varied (i.e., density, branch number, or CDT CSA). The presentation of our results follows this structure. 

\section{Results}\label{sec:Results Mesh Density}

\subsection{Branched vs. Non-Branched CDT}

\begin{figure}[H]
\includegraphics[width=0.6\textwidth]{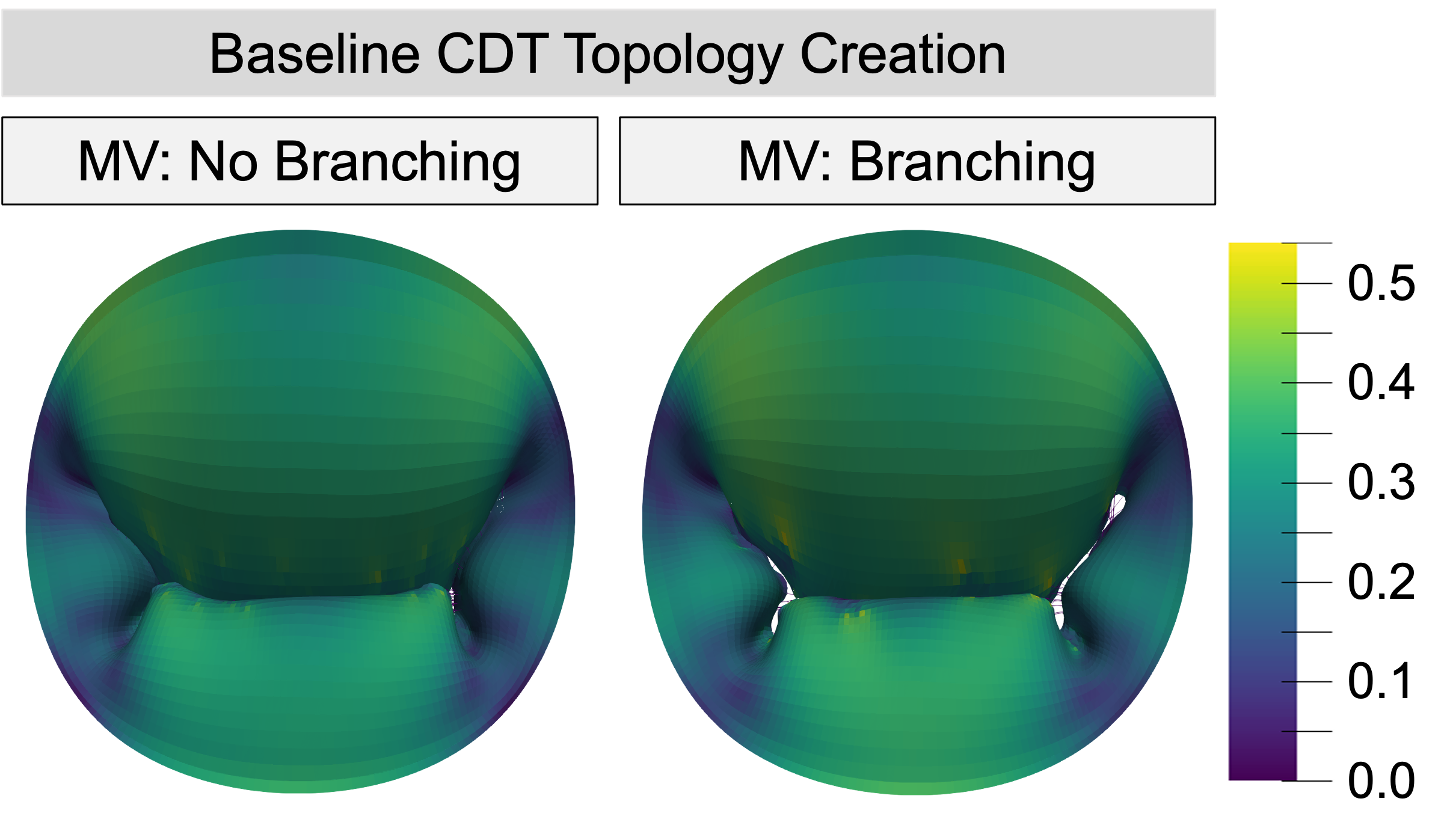}
\centering
\caption{Lagrange strain color maps comparing a MV with non branching spring elements and 114 insertion points to our novel CDT creation method with branching linear truss elements and 115 insertion points.}
\label{Figure: Baseline Topology}
\end{figure}

Despite having a similar number of insertion points, transitioning from the non-branched spring CDT to the truss models caused the valve closure to change. The maximum strain and ROA increased by $26\%$ and $135\%$, respectively, while the CA and billowing volume decreased by $4.8\%$ and $25\%$ (Fig.~\ref{Figure: Baseline Topology}).

\subsection{Mesh density}

\begin{figure}[H]
\includegraphics[width=0.30\textwidth]{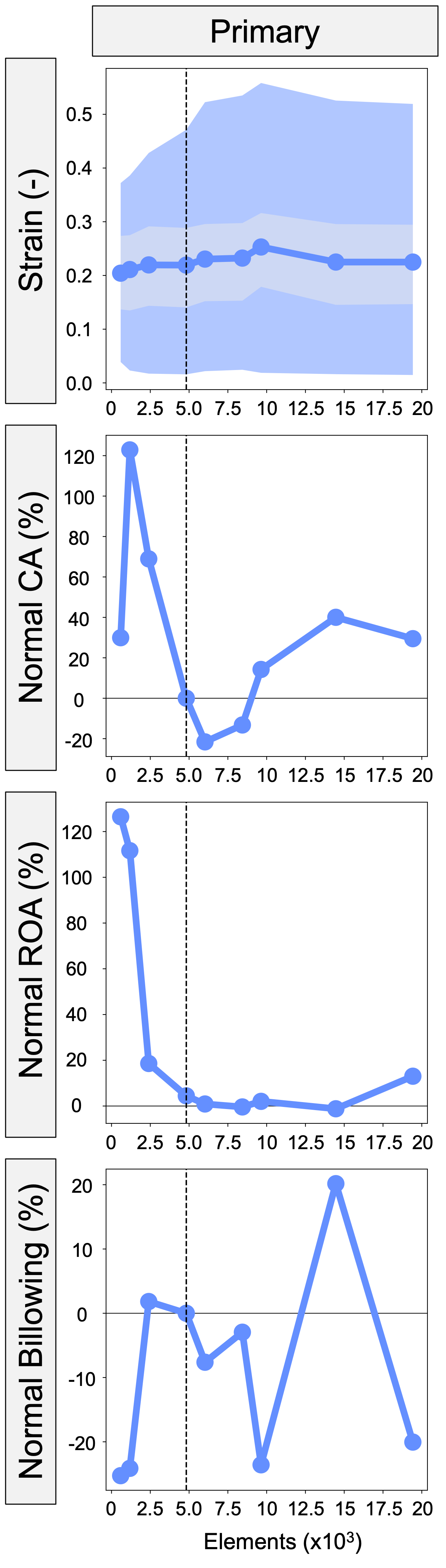}
\centering
\caption{Different FE mesh sizes were evaluated in relation to the valve closure metrics (strain, ROA, CA, billowing volume) with a 4820-element mesh chosen for the parametric variation study.}
\label{Figure: Mesh Density Study}
\end{figure}

Results of the mesh convergence study are shown in Figure~\ref{Figure: Mesh Density Study}. A finite element mesh with $4820$ elements was chosen for the parametric variation study presented in Sections~\ref{sec:ResultsPrimaryCT} -- \ref{sec:ResultsAddValves} given this mesh density minimized simulation times to $227$\,seconds while still meeting criteria for FE model convergence of ROA with minimal variation in the strain metrics. Fluctuations in the CA and billowing volume beyond this mesh density may stem from localized changes in leaflet curvature that would influence these values; however, this effect should be minimized in our study results since the same mesh density was used throughout.

\subsection{Primary CDT Geometry}
\label{sec:ResultsPrimaryCT}
\begin{figure}[h]
\includegraphics[width=0.95\textwidth]{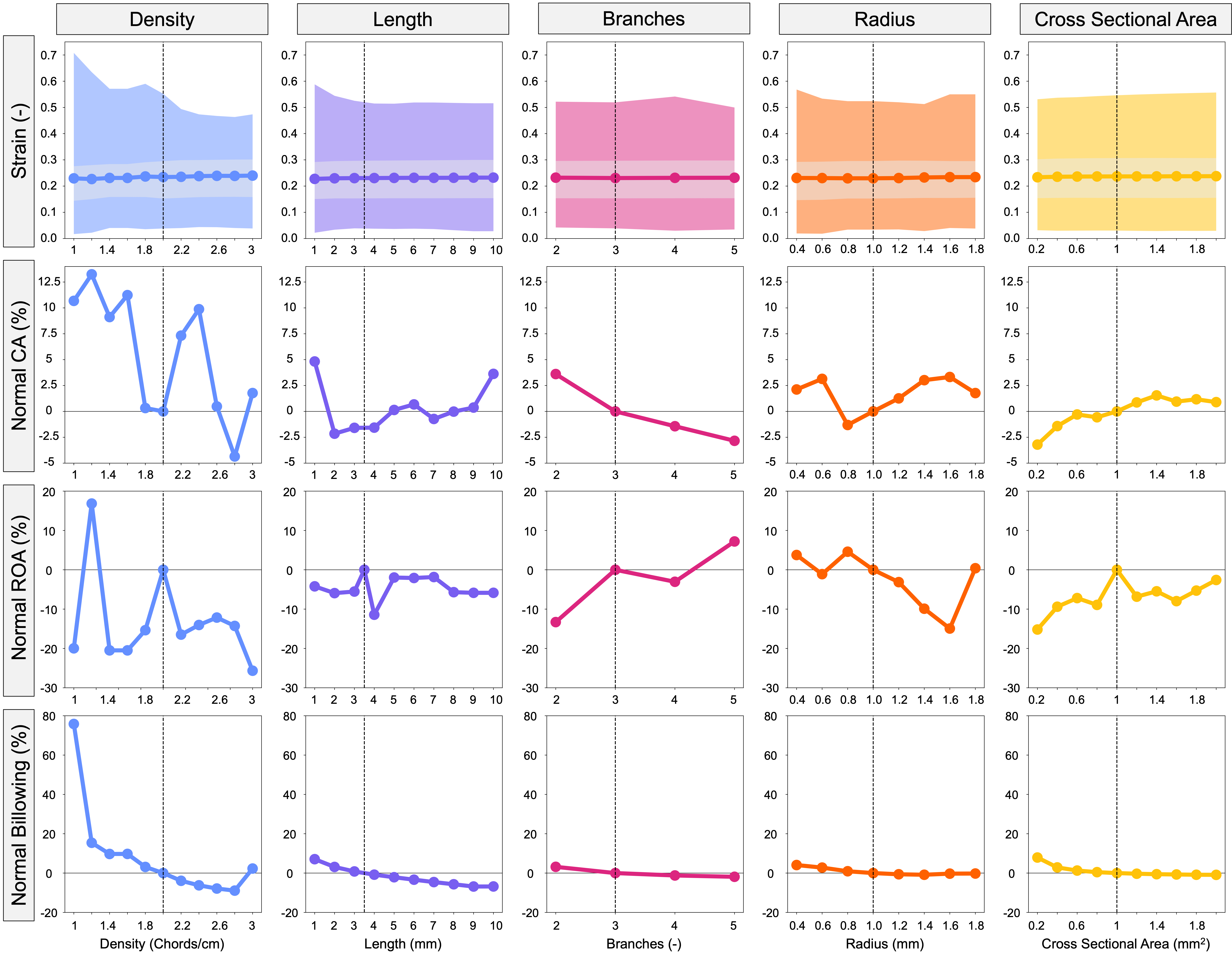}
\centering
\caption{Influence of primary CDT parameters (density, length, branches, radius, and CSA) on valve closure metrics. The shaded regions in the strain plots represent the maximum (darker) and inter quartile range (lighter). CA, ROA, and billowing volume were normalized to the MV with baseline CDT topology indicated with a vertical dashed line.}
\label{Figure: Primary CT Study}
\end{figure}

We first altered the primary CDT parameters to examine their effect on valve closure (Fig.~\ref{Figure: Primary CT Study}). 

\subsubsection{Density}
As the primary CDT density increases, the maximum strain decreases from $1$\,${\rm chords}/{\rm cm}$ ($0.71$) to $3$\,${\rm chords}/{\rm cm}$ ($0.47$). Variation in density resulted in contact area reaching a maximum at $1.2$\,${\rm chords}/{\rm cm}$ ($85.78$\,${\rm mm}^2$ -- $13.24\%$ above baseline), and a minimum at $2.8$\,${\rm chords}/{\rm cm}$ ($73.76$\,${\rm mm}^2$ -- $4.36\%$ below baseline), yet having no discernible trend. No trend was observed with the ROA either, but values varied between extrema of $1.2$\,${\rm chords}/{\rm cm}$ ($29.14$\,${\rm mm}^2$ -- $16.84\%$ above baseline) to $3$\,${\rm chords}/{\rm cm}$ ($18.54$\,${\rm mm}^2$ -- $25.66\%$ below baseline). However, the billowing volume generally decreased monotonically from $1$\,${\rm chords}/{\rm cm}$ ($1214.2$\,${\rm mm}^3$ -- $75\%$ above baseline) to $3$\,${\rm chords}/{\rm cm}$ ($706.5$\,${\rm mm}^3$ -- $2.3\%$ below baseline).

\subsubsection{Length}
A well-defined trend for CA and ROA vs. primary CDT length was not observed. However, maximum strain decreased from from $1$\,mm ($0.59$) to $10$\,mm ($0.52$). Billowing volume also decreased from $1$\,mm ($739.1$\,${\rm mm}^3$ -- $7.1\%$ above baseline) to $10$\,mm ($643.3$\,${\rm mm}^3$ -- $6.8\%$ below baseline).

\subsubsection{Number of Branches}
The valve closure metrics generally did not exhibit a trend vs. number of primary CDT branches except in relation to CA, as CA decreases from $2$\,branches ($78.89$\,${\rm mm}^2$ -- $3.6\%$ above baseline) to $5$\,branches ($73.98$\,${\rm mm}^2$ -- $2.84\%$ below baseline). 

\subsubsection{Branch Radius}
Strain and CA exhibited mild sensitivity to radius variation, and this parameter was able to influence ROA and billowing volume, but in general, no trends were observed. ROA reached a maximum at $0.8$\,mm ($23.67$\,${\rm mm}^2$ -- $4.6\%$ above baseline) and a minimum at $1.6$\,mm ($19.25$\,${\rm mm}^2$ -- $14.93\%$ below baseline), with the billowing volume slightly decreasing from $0.4$\,mm ($718$\,${\rm mm}^3$ -- $4.09\%$ above baseline) to $1.8$\,mm ($689$\,${\rm mm}^3$ -- $0.23\%$ below baseline). 

\subsubsection{Cross-Sectional Area}
Increasing CSA from $0.02$ to $0.20$\,${\rm mm}^2$ demonstrated influence over all valve closure metrics. Maximum strain increased from $0.2$\,${\rm mm}^2$ ($0.53$) to $2.0$\,${\rm mm}^2$ ($0.56$). The CA increased from $0.2$\,${\rm mm}^2$ ($75.78$\,${\rm mm}^2$ -- $3.21\%$ below baseline) to $2.0$\,${\rm mm}^2$ ($78.99$\,${\rm mm}^2$ -- $0.9\%$ above baseline). The ROA reached a minimum at $0.2$\,${\rm mm}^2$ ($23.57$\,${\rm mm}^2$ -- $15\%$ below baseline) and a maximum at $1.0$\,${\rm mm}^2$ ($27.79$\,${\rm mm}^2$ -- at baseline), but with no discernible trend. Billowing volume decreased from $0.2$\,${\rm mm}^2$ ($622.3$\,${\rm mm}^3$ -- $7.91\%$ above baseline) to $2.0$\,mm ($689$\,${\rm mm}^3$ -- $0.23\%$ below baseline).

\subsection{Secondary CDT Geometry}
\begin{figure}[h]
\includegraphics[width=0.95\textwidth]{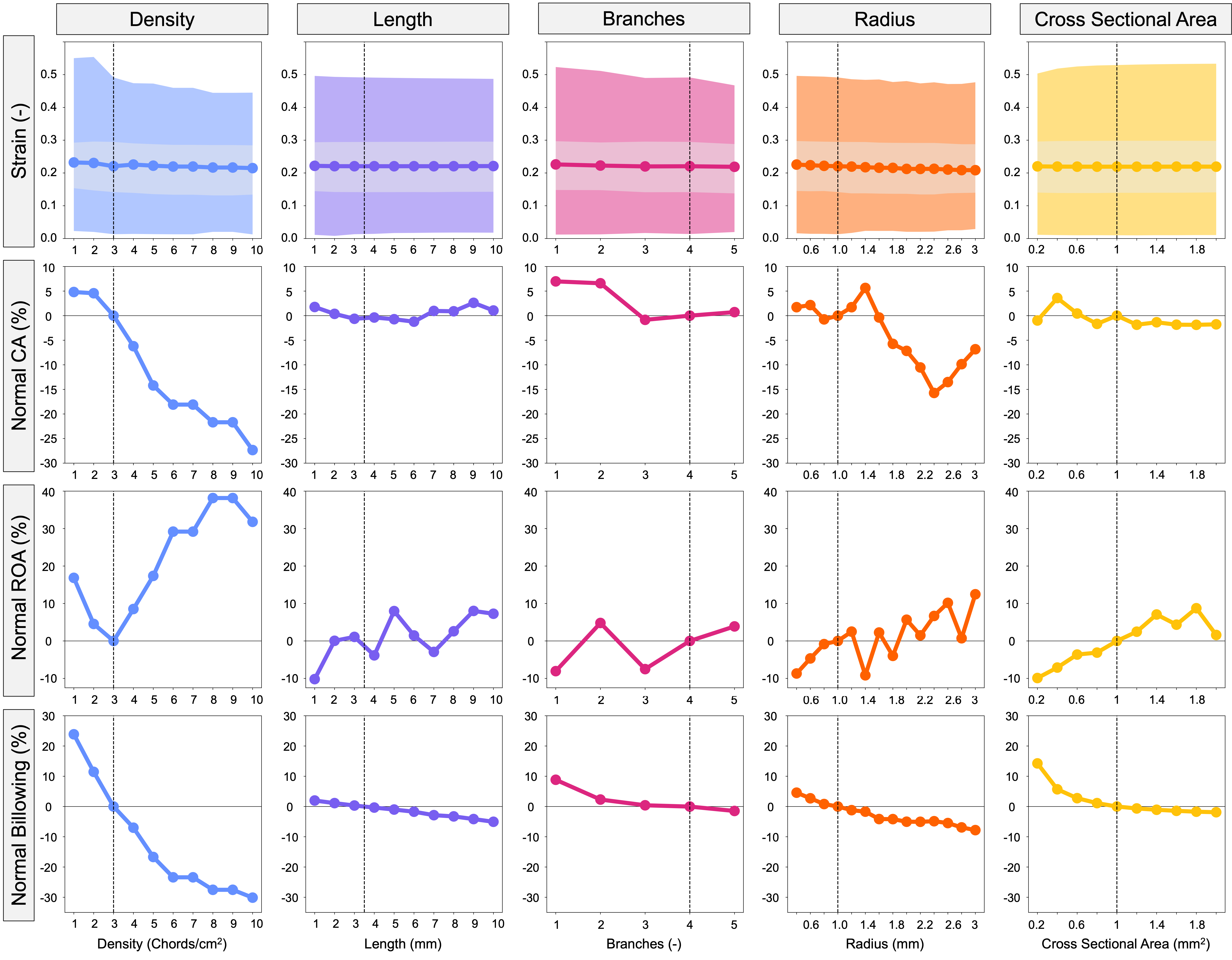}
\centering
\caption{Influence of secondary CDT parameters (density, length, branches, radius, and CSA) on valve closure metrics. The shaded regions in the strain plots represent the max- and inter-quartile range. CA, ROA, and billowing volume were normalized to the MV with baseline CDT topology indicated with a vertical dashed line.}
\label{Figure: Secondary CT Study}
\end{figure}

We also investigated how secondary CDT parameters influence valve closure metrics (Figure~\ref{Figure: Secondary CT Study}). In many ways the secondary CDT metrics reflected the findings of the primary CDT (Section~\ref{sec:ResultsPrimaryCT}). 

\subsubsection{Density}
Increasing the secondary CDT density from $1$\,${\rm chords/cm}^2$ to $10$\,${\rm chords/cm}^2$ resulted in the maximum first principal strain decreasing from $0.55$ to $0.45$. However, in contrast to the primary CDT study, increasing the secondary CDT density resulted in a stronger influence over CA, ROA, and billowing volume. The CA decreased from $1$\,${\rm chords/cm}^2$ ($72.78$\,${\rm mm}^2$ -- $4.81\%$ above baseline) to $10$\,${\rm chords/cm}^2$ ($50.46$\,${\rm mm}^2$ -- $27.34\%$ below baseline). ROA moderately varied in relation to secondary CDT density with the minimum density model exhibiting a $16.85$\% increase above baseline ($29.06$\,${\rm mm}^2$), with ROA decreasing from there to the baseline value ($24.87$\,${\rm mm}^2$) at $3$\,${\rm chords/cm}^2$ (similar to the primary CDT value). But in contrast to the primary CDT variation behavior, the parameter increased from baseline by $31.81$\% (32.768\,${\rm mm}^2$) for the maximum density model. 

\subsubsection{Length}
Variation of the secondary CDT length produced similar results to the primary study as they both influence strain and billowing volume, and do not influence CA. ROA however, increased from $1$\,mm ($21.51$\,${\rm mm}^2$ -- $10.19\%$ below baseline) to $10$\,mm ($25.69$\,${\rm mm}^2$ -- $7.27\%$ above baseline) with non-monotonic variation between extrema.  

\subsubsection{Number of Branches}
Results due to increasing the number of secondary CDT branches did not demonstrate any similarities with the primary CDT study. Here, the maximum strain decreased from $1$\,branches ($0.52$) to $5$\,branches ($0.47$). Additionally, billowing decreased from $1$\,branches ($533$\,${\rm mm}^3$ -- $8.86\%$ above baseline) to $5$\,branches ($482.5$\,${\rm mm}^3$ -- $4.09\%$ below baseline).

\subsubsection{Branch Radius}
Secondary CDT radius demonstrated more influence over all valve closure metrics than in primary CDT radius variation. Maximum strain decreased from $0.4$\,mm ($0.50$) to $3$\,mm ($0.48$). Contact area peaked at $1.4$\,mm ($73.36$\,${\rm mm}^2$ -- $5.63\%$ above baseline) and had a minimum at $2.4$\,mm ($58.53$\,${\rm mm}^2$ -- $15.71\%$ below baseline), but with no apparent trend. ROA generally increased with secondary radius, with a minimum at $0.4$\,mm ($22.71$\,${\rm mm}^2$ -- $8.69\%$ below baseline) and a maximum at $3$\,mm ($27.97$\,${\rm mm}^2$ -- $12.46\%$ above baseline). The billowing volume decreased from $0.4$\,mm ($512.1$\,${\rm mm}^3$ -- $4.57\%$ above baseline) to $3$\,mm ($451.5$\,${\rm mm}^3$ -- $7.8\%$ below baseline).

\subsubsection{Cross-Sectional Area}
The secondary CDT CSA variation produced similar results to those observed in the primary CDT CSA study as it demonstrated influence over strain, ROA, and billowing volume, with the exception of it exhibiting only trivial sensitivity on CA. The maximum strain increased from $0.2$\,${\rm mm}^2$ ($0.50$) to $2.0$\,${\rm mm}^2$ ($0.53$). The ROA increased from $0.2$\,${\rm mm}^2$ ($22.58$\,${\rm mm}^2$ -- $9.90\%$ below baseline) to $2.0$\,${\rm mm}^2$ ($25.46$\,${\rm mm}^2$ -- $1.6\%$ above baseline). The billowing volume decreased from $0.2$\,${\rm mm}^2$ ($447.3$\,${\rm mm}^3$ -- $14.25\%$ above baseline) to $2.0$\,${\rm mm}^2$ ($384.1$\,${\rm mm}^3$ -- $1.89\%$ below baseline).

\subsection{Insertion Points and Total CSA}
\begin{figure}[h]
\includegraphics[width=0.5\textwidth]{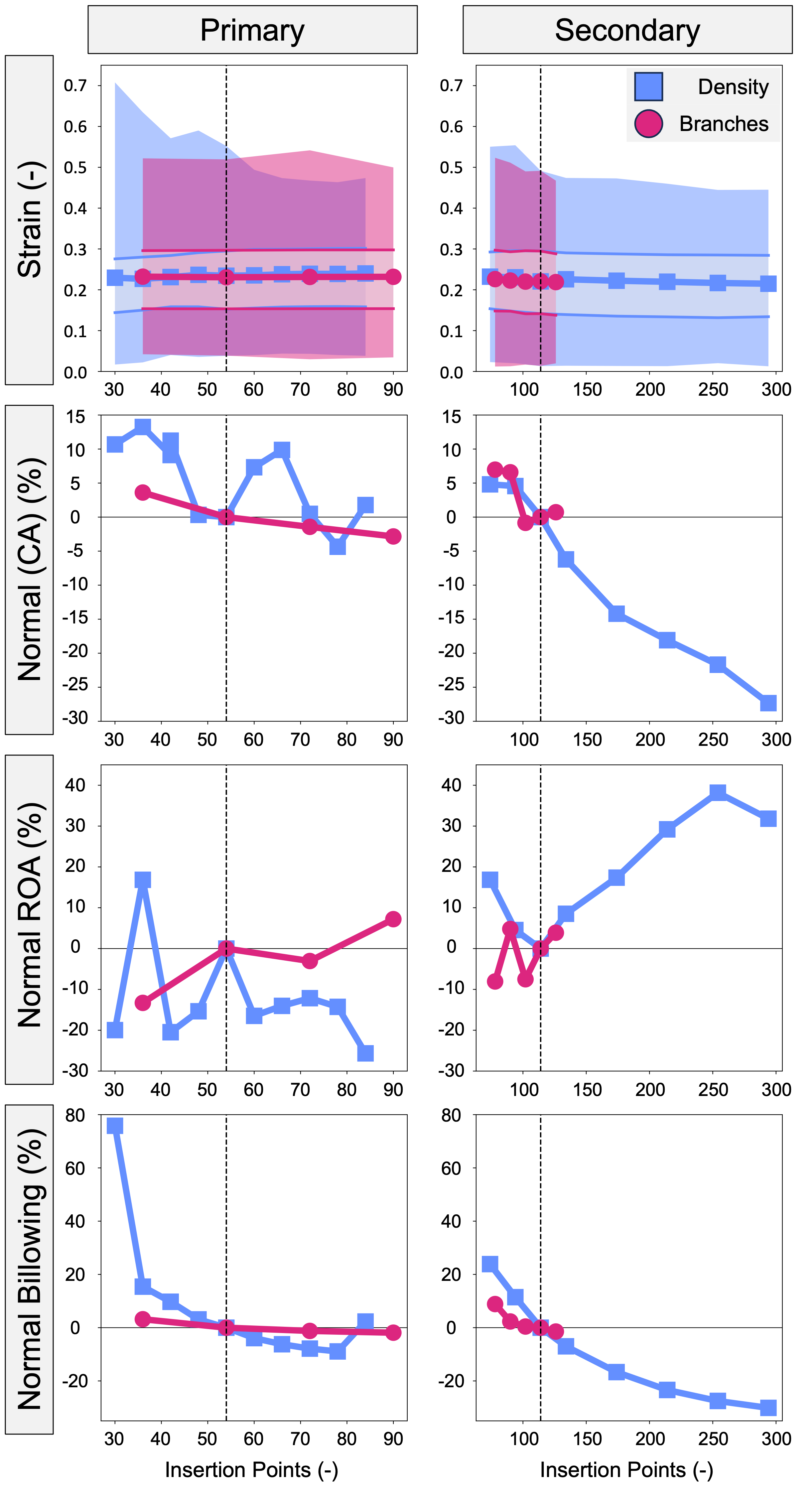}
\centering
\caption{The effect of density and branches on valve closure metrics when evaluated by the number of insertion points. The shaded regions in the strain plots represent the maximum (darker) and inter quartile range (lighter).}
\label{Figure: Insertion Points}
\end{figure}

The CDT density and branching metrics were re-quantified as a function of total CDT insertion points and are presented in Figure~\ref{Figure: Insertion Points}. Interestingly, there were no consistent trends when varying the density and number of branches for the primary CDT, suggesting these parameters independently influence valve function. In contrast, there were decreasing trends in maximum strain, median strain, and billowing volume when secondary CDT density and number of branches were increased. CDT density appeared to have a stronger influence on these metrics, but this comparison is inhibited by the relatively limited range when varying the number of branches. 

\begin{figure}[h]
\includegraphics[width=0.5\textwidth]{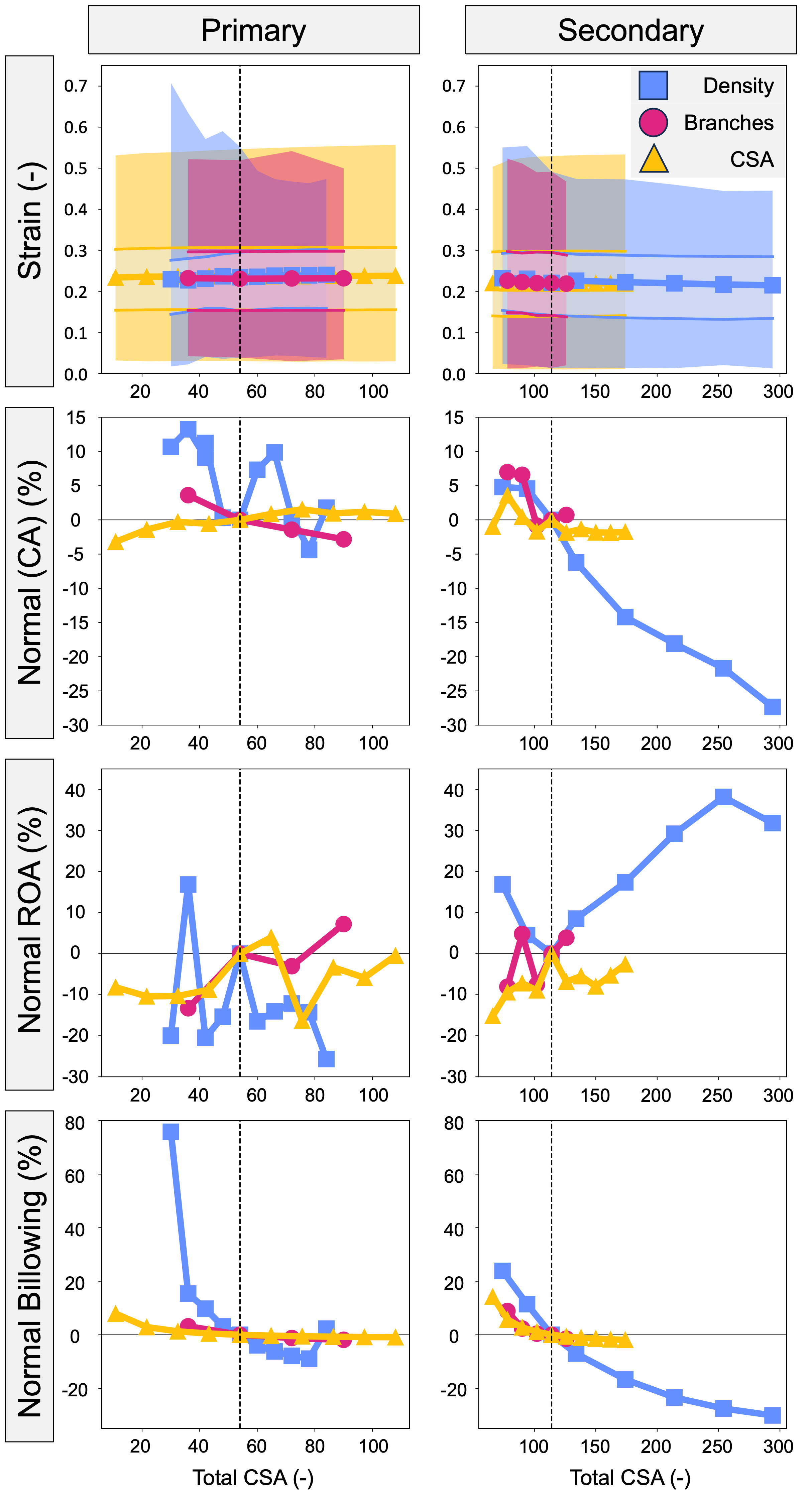}
\centering
\caption{The effect of density, branches, and CSA on valve closure metrics based on the total CSA of the CDT. The shaded regions in the strain plots represent the maximum (darker) and inter quartile range (lighter).}
\label{Figure: TCSA}
\end{figure}

Additionally, the CDT density, branching, and CSA metrics were re-quantified as a function of total CDT CSA and are presented in Figure~\ref{Figure: TCSA}. Similar to our analysis based on CDT insertion point density, there were no consistent trends when varying the density, branches, or CSA of the primary CDT. However, there were consistent decreases in median strain and billowing volume as the total secondary CDT CSA was increased, suggesting total secondary CDT CSA is a crucial consideration to simulations of valve function. 

\subsection{Additional Valves}
\label{sec:ResultsAddValves}
\begin{figure}[h]
\includegraphics[width=0.6\textwidth]{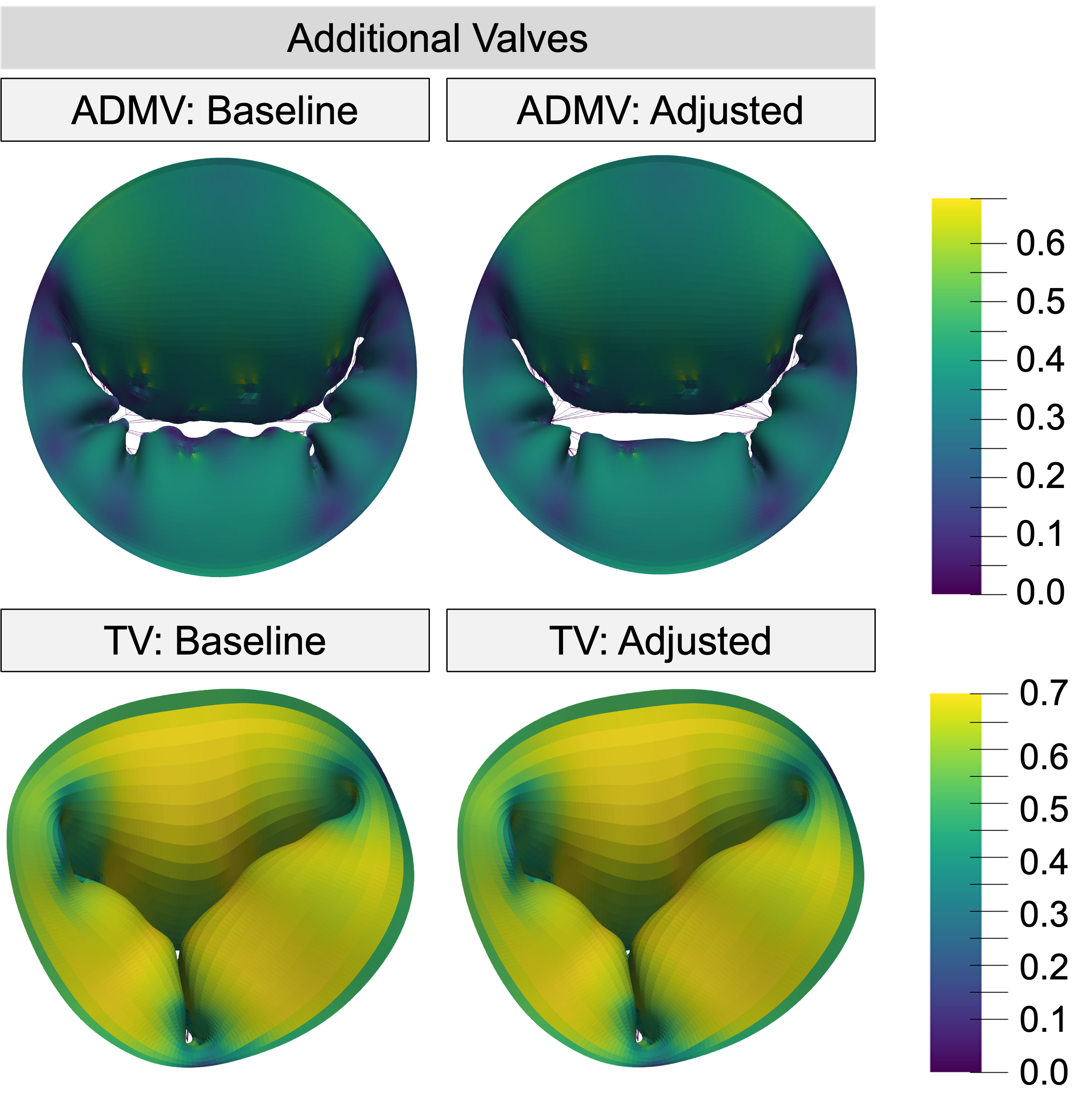}
\centering
\caption{Strain color maps comparing a MV with non branching spring elements and 114 insertion points to our novel CDT creation method with branching linear truss elements and 115 insertion points.}
\label{Figure: Additional Valves}
\end{figure}

To demonstrate the utility of our new parametric CDT generation tool, the CDT geometries of the ADMV and TV were successfully adjusted from the baseline CDT to improve simulated valve function (Fig.~\ref{Figure: Additional Valves}). The primary CDT density of the ADMV model was increased to reduce local free edge prolapse, which led to a $13.5$\% decrease in CA and $17.4$\% increase in ROA. Additionally, the secondary CDT density of the baseline TV model was decreased to reduce the ROA by $13$\% and better resemble a functioning, stereotypical TV. Decreasing the secondary CDT also resulted in a minimal change in maximum strain ($0.629$ -- $0.628$), decreased CA ($4.84$\%), and increased billowing volume ($5.56$\%). These extensions of our pipeline illustrate how the CDT generation tool can be successfully used for a range of AVV geometries. 

\section{Discussion}\label{sec:Discussion}

\subsection{Overall Findings}
In this study, we investigated how variations in CDT structure influence AVV closure by combining our new open-source tools in SlicerHeart with our established FEBio workflow (e.g.,~\cite{wu2022computational, wu2023effects}). Our findings demonstrated that altering the CDT geometry of a stereotypical MV led to changes in clinically relevant valve metrics (ROA, CA, and billowing volume) and valve mechanics (first principal strains). Specifically, CSA had the most influence over valve closure metrics, followed by CDT density, length, radius and branches. We then used this information to showcase the flexibility of our new workflow by altering the CDT geometry of two additional valve topologies (MV with annular dilation and TV) to improve FE predictions. To our knowledge, this is the first study to generate morphologically realistic CDT geometries and parametrically investigate how CDT branching influences FE predictions of AVV function.  

\subsection{Existing Literature}

There has been great progress in computational predictions of AVV closure over the past two decades~\cite{stevanella2010finite, wang2013finite, kong2018finite, khalighi2019development, mathur2020tricuspid, laurence2020pilot, johnson2021parameterization, sturla2014possible, choi2014novel, kong2018repair, laurence2020pilot, kong2020finite}, yet accurate reconstruction of \textit{in vivo} CDT geometries remains an unresolved challenge. This has led to research groups adopting various methods (see Table~\ref{Table:ExistingLiterature}) in which the CDT geometries are: (i) accurately reconstructed from \textit{ex vivo} samples, (ii) informed from clinical images and numerical optimization, (iii) artificially generated based on assumed CDT properties, or (iv) completely disregarded or simplified to force or displacement boundary conditions acting on the leaflets. Here, we review each classification of CDT geometry fidelity to contextualize our newly developed CDT generation tool. 

\begin{sidewaystable}[]
\caption{Previous literature on computational modeling of AVVs grouped by CDT implementation in FE modeling. Abbreviations: CDT - Chordae Tendineae, Img. Der. - Image Derived, Img. Inf. - Image Informed, Synth. - Synthetic, Y - Yes, N - No, AL - Anterior Leaflet, PL - Posterior Leaflet, SL - Septal Leaflet, BSM - Basal, Strut, Marginal, Het. - Heterogeneous, CSA - Cross-Sectional Area, MV - Mitral Valve, TV - Tricuspid Valve.}
\label{Table:ExistingLiterature}
\begin{tabular*}{\textwidth}{@{\extracolsep{\fill}}lccccccc}
\toprule
\textbf{Publication}   & \textbf{\begin{tabular}[c]{@{}c@{}}CDT\\ Fidelity\end{tabular}} & \textbf{\begin{tabular}[c]{@{}c@{}}CDT Insertion\\ Classification\end{tabular}} & \textbf{\begin{tabular}[c]{@{}c@{}}Branching\\ Points\end{tabular}} & \textbf{\begin{tabular}[c]{@{}c@{}}CSA\\ (${\rm mm}^2$)\end{tabular}} & \textbf{\begin{tabular}[c]{@{}c@{}}CDT\\ Origins\end{tabular}} & \textbf{\begin{tabular}[c]{@{}c@{}}CDT\\ Arc\end{tabular}} & \textbf{Valve} \\ \midrule
Drach \textit{et al}. 2018~\cite{drach2018comprehensive}      & Img. Der. & Y & Het. & Het. & 12$^\dagger$ & Y & \textit{Ex Vivo} MV \\
Mathur \textit{et al}. 2022~\cite{mathur2022texas}     & Img. Der. & AL, PL, SL & 1 & 0.33, 0.32, 0.40 & 4 & Y & \textit{Ex Vivo} TV \\
Laurence \textit{et al}. 2020~\cite{laurence2020pilot}   & Img. Der. & Y & 3 & 0.17 & 3 & Y & \textit{Ex Vivo} TV \\
Wang \textit{et al}. 2013~\cite{wang2013finite}       & Img. Inf. & BSM & 1 & 0.71, 2.05, 0.38 & 21 & N & \textit{In Vivo} MV \\
Pham \textit{et al}. 2017~\cite{pham2017finite}       & Img. Inf. & BSM & 1 & 0.71, 2.05, 0.38 & 18 & Y & \textit{In Vivo} MV \\
Kong \textit{et al}. 2018~\cite{kong2018finite}       & Img. Inf. & Y & 1 & 0.95 & $\dagger$ & Y & \textit{In Vivo} TV \\
Khalighi \textit{et al}. 2019~\cite{khalighi2019development}   & Synth. & N & 0 & Constant$^\dagger$ & 2 & N & \textit{Ex Vivo} MV \\
Johnson \textit{et al}. 2021~\cite{johnson2021parameterization}    & Synth. & Y & 0 & 0.166 & 4 & Y & Parametric  TV \\
Stevanella \textit{et al}. 2010~\cite{stevanella2010finite} & Synth. & Y & 0 & 0.171 & 3 & Y & \textit{In Vivo} TV \\
Wenk \textit{et al}. 2010~\cite{wenk2010first}       & Synth. & Y & 0 & Constant$^\dagger$ & 4$^\dagger$ & N & \textit{In Vivo} MV \\
Wu \textit{et al}. 2023~\cite{wu2023effects}         & Synth. & Y & 0 & N/A & 3 & N & \textit{In Vivo} TV \\ \bottomrule
\multicolumn{8}{l}{$^\dagger$Not reported. Note: Values were interpreted from figures where possible.} 
\end{tabular*}
\end{sidewaystable}

\subsubsection{High-Fidelity Chordae Reconstruction}
\textit{Ex vivo} methods are currently the only option to fully capture complex CDT geometry. For example, Drach \textit{et al}.~\cite{drach2018comprehensive} generated micro-CT scans of an explanted ovine MV to fully reconstruct the valve, annulus, and CDT in both unloaded/loaded configurations. The full reconstruction of the CDT origins, branching, and CSA allowed accurate predictions of the MV closure and mechanics in healthy, diseased, and repaired configurations. Explanted valves also enable researchers to measure, count, and virtually reconstruct CDT geometry in the absence of high-fidelity imaging~\cite{mathur2022texas, laurence2020pilot}, although this method may be more prone to user error and bias. In general, the main challenge precluding these methods for clinical application is the need of an explanted valve; however, photon-counting CT is an emerging clinical imaging modality that may allow for accurate virtual reconstruction of \textit{in vivo} CDT structure~\cite{douek2023clinical}. 

\subsubsection{Image-Informed Chordae Reconstruction}
To partially overcome the challenges of obtaining complete \textit{in vivo} CDT geometries, some groups have used the partial visualization of CDT geometries in clinical images to inform CDT geometry and function. Most notably, this has been achieved with multi-slice CT scans that can resolve the PM, CDT origins, and some degree of CDT topology and leaflet insertion points~\cite{wang2013finite, pham2017finite, kong2018finite}. This information is used to infer the CDT geometry with critical parameters (e.g., length) optimized by comparing FE predictions of valve closure to the image-derived leaflet surface. Although this approach can successfully match \textit{in vivo} valve behavior, it is possible that patient-specific leaflet behaviors or other simulation discrepancies are implicitly embedded in the optimized CDT parameter(s). 

\subsubsection{Purely Artificial Chordae Construction}
In contrast to image-informed CDT reconstruction, purely artificial methods disregard any information possibly inferred from clinical images and synthetically construct CDT geometries. These efforts were largely motivated by the work of Khalighi \textit{et al}.~\cite{khalighi2019development} which iteratively simplified a high-resolution micro-CT scan of an ovine heart (see also~\cite{drach2018comprehensive}) to understand the minimum required CDT fidelity to emulate CDT function. Their foundational work demonstrated that the CDT structure can be significantly simplified to have (i) no branches, (ii) constant CSA, (iii) combined CDT origins (i.e., PMs), and (iv) random leaflet insertions so long as the insertion density was above a critical value. Later works by our group~\cite{wu2022computational, wu2023effects} used this when developing our open-source FEBio workflow for heart valves. Other groups established their own methods for parametrically defining chordae structure~\cite{johnson2021parameterization} or creating artificial CDT geometries~\cite{stevanella2010finite}.

\subsubsection{Emulating Chordae Behavior}
A  sub-class of FE simulations focused on matching predicted leaflet closure to clinical images have entirely removed the CDT geometry and emulated CDT behavior via force or displacement boundary conditions. The FE shape enforcement work of Rego \textit{et al}.~\cite{rego2018noninvasive} compared forced-based methods for emulating chordae behavior and found similar performance whether applied to the entire leaflet, the leaflet rough zone, or the leaflet free edge. We used these findings to remove CDT from our recent development of an open-source FE shape enforcement method in FEBio~\cite{laurence2024}. Alternatively, Ross \textit{et al}.~\cite{ross2024bayesian} demonstrated a method to directly prescribe leaflet free edge motion from clinical images which substantially improved optimization search results when estimating \textit{in vivo} leaflet properties. This method performs best when the FE prediction is being matched to an image-derived leaflet surface via shape enforcement or material optimization, but a chordae emulating force needs careful tuning to ensure no unnecessary tensions are applied to the leaflets. Furthermore, this method cannot be used when altering the valve geometry (i.e., simulating diseased-driven changes or surgical repair), limiting clinical application only to estimation of \textit{in vivo} leaflet strains or material properties.  

\subsubsection{The Present Study}

The workflow presented herein has expanded upon purely artificial methods for reconstructing CDT geometry by including CDT branching with precise parametric control over the branching properties (i.e., branch length and branch number). We have used our open-source CDT generation tool to provide the first parametric assessment of how CDT geometry influences AVV closure, which will support future FE simulations of AVV function. The current framework also serves as a valuable foundation for generating CDT geometries with more morphological considerations (e.g., varying CSA) or CDT geometries informed by clinical images. Therefore, we have intentionally developed our CDT generation tools open-source in the SlicerHeart extension for 3D Slicer to (i) provide an open-source foundation for other research groups to expand upon our parametric generation of CDT geometries and (ii) allow smooth integration with clinical imaging data and the open-source FEBio software. 

Certain CDT parameters had more influence over valve closure metrics than others. CDT density consistently had some of the largest variations from baseline values. This may be due to the fact that the density parameter causes a restructuring of the CDT topology while parameters such as number of branches and branching radius which only have a localized effect restricted to the insertion sight. The large variations caused by the density parameter may also be do to the fact that material property changes such as total CSA can have large changes with the changing number of CDT. The restructuring of the CDT topology by changing the density can specifically have a large impact on the commissure folding which can influence strain and CA. 

\subsection{Toward Clinical Application}
Realistic modeling of chordal structures is important to future simulations of valve function, valve dysfunction, and valve repair. For example, alteration of chordal structure, such as the introduction of ``neo-chords" is an important part of many clinically applied valve repairs~\cite{elde2023neochords}. A significant question in such repairs is how many and how long should the neo-chords be, and modeling of such interventions requiring modeling of CDT topologies beyond the simple, non-branching, functionally-equivalent frameworks demonstrated in Khalighi \textit{et al}.~\cite{khalighi2019development}. In contrast, our framework allows the reproducible creation and refinement of more realistic chordal branching which allows simulation of these more complex chordal interventions. Currently, accurate ``subject-specific" chordal models can only reliably be obtained from direct inspection or micro-ct imaging of static \textit{ex vivo} models~\cite{sacks2019simulation}. However, emerging technology such as photon-counting CT may provide the fidelity to accurately visualize the majority of the chordal structure~\cite{douek2023clinical}. As such technology evolves, greater fidelity of chordal modeling can be incorporated into patient-specific studies.

\subsection{Study Limitations}
The study completed above sought to incorporate all information currently available in the literature; however, simplifying assumptions were still required. The topological models of the AVVs were approximated by merging \textit{in vivo} imaging with \textit{ex vivo} data reported in the literature. Future work will include patient-specific models from \textit{in vivo} 3DE imaging. The use of patient-specific imaging will also allow us to address the dynamics of the annulus that we currently assumed as pinned in this study. In addition, actual PM locations can be incorporated from individual patient images.

In addition to the geometry, the leaflets were simplified with a homogeneous thickness taken from \textit{ex vivo} data. Similarly, the leaflet constitutive model was taken from \textit{ex vivo} mechanical testing as there is no current method to extract patient-specific material properties from \textit{in vivo} 3DE imaging. Our group and others are developing \textit{in-silico} tools for these purposes~\cite{wu2024adept, ross2024bayesian}. We used a linear material model as a simplifying assumption for our CDT, but studies have shown that the Ogden material model would better represent CDT tissue~\cite{smith2021tricuspid}.

Finally, commissure folding is challenging to model and slight alterations to the CDT geometry may cause relatively large variations in the localized commissural buckling. We believe this may contribute to the fluctuations in contact area throughout our results. In the future, we can extend our methods for image-informed biomechanics (e.g.,~\cite{laurence2024, wu2024adept}) to improve simulation performance in these valve regions.

\backmatter

\bmhead{Declaration of Conflicts of Interest}

The authors of this paper have no financial or personal relationships with other people or organizations that could inappropriately influence/bias this work.

\bmhead{Supplementary information}

Please find Video S1 attached as \texttt{VideoS1.mp4}.

\bmhead{Acknowledgments}

Supported by The Cora Topolewski Pediatric Valve Center at CHOP, an Additional Ventures Single Ventricle Research Fund Grant, NIH R01 HL153166, R01 GM083925 (SAM, JAW), K25 HL168235 (WW), and T32 HL007915 (DWL).

\newpage

\begin{appendices}

\section{Tabulated metrics}\label{secA1}

The first principal strain values for the CDT parameters (branches, CSA, density, and radius) are provided in Tables~\ref{Table: Strain Branches}-\ref{Table: Strain Length}.

\begin{sidewaystable}[]
\caption{First principle strain values are reported here for each primary and secondary parameter of branching CDT. The strains are reported at various percentiles (maximum, 99th, 95th, 75th, 25th, 5th, 1st, and minimum) as well as the median strain.}
\label{Table: Strain Branches}
\begin{tabular}{ccccccccccc}
\hline
{\textbf{CDT Classification}} & {\textbf{Branches (-)}} & \textbf{max} & \textbf{99\%} & \textbf{95\%} & \textbf{75\%} & \textbf{median} & \textbf{25\%} & \textbf{5\%} & \textbf{1\%} & \textbf{min} \\ \hline
& 2 & 0.52 & 0.41 & 0.35 & 0.30 & 0.23 & 0.15 & 0.09 & 0.06 & 0.04 \\
& 3 & 0.52 & 0.40 & 0.35 & 0.30 & 0.23 & 0.15 & 0.09 & 0.06 & 0.04 \\
& 4 & 0.54 & 0.40 & 0.35 & 0.30 & 0.23 & 0.15 & 0.09 & 0.06 & 0.03 \\
\multirow{-4}{*}{Primary} & 5 & 0.50 & 0.41 & 0.35 & 0.30 & 0.23 & 0.15 & 0.09 & 0.06 & 0.03 \\ \hline
& 1 & 0.52 & 0.37 & 0.34 & 0.30 & 0.23 & 0.15 & 0.09 & 0.05 & 0.01 \\
& 2 & 0.51 & 0.37 & 0.34 & 0.29 & 0.22 & 0.15 & 0.08 & 0.05 & 0.01 \\
& 3 & 0.49 & 0.36 & 0.34 & 0.30 & 0.22 & 0.14 & 0.08 & 0.05 & 0.02 \\
& 4 & 0.49 & 0.36 & 0.34 & 0.29 & 0.22 & 0.14 & 0.08 & 0.05 & 0.01 \\
\multirow{-5}{*}{Secondary} & 5 & 0.47 & 0.36 & 0.33 & 0.29 & 0.22 & 0.14 & 0.08 & 0.05 & 0.02 \\ \hline
\end{tabular}
\end{sidewaystable}

\begin{sidewaystable}[]
\caption{First principle strain values are reported here for each primary and secondary parameter of varying CSA. The strains are reported at various percentiles (maximum, 99th, 95th, 75th, 25th, 5th, 1st, and minimum) as well as the median strain.}
\label{Table: Strain CSA}
\begin{tabular}{ccccccccccc}
\hline
\textbf{CDT Classification} & \textbf{CSA ${\rm mm}^2$} & \textbf{max} & \textbf{99\%} & \textbf{95\%} & \textbf{75\%} & \textbf{median} & \textbf{25\%} & \textbf{5\%} & \textbf{1\%} & \textbf{min} \\ \hline
\multirow{10}{*}{Primary} & 0.2 & 0.53 & 0.41 & 0.35 & 0.30 & 0.23 & 0.15 & 0.09 & 0.06 & 0.03 \\
 & 0.4 & 0.54 & 0.41 & 0.36 & 0.30 & 0.24 & 0.15 & 0.09 & 0.06 & 0.03 \\
 & 0.6 & 0.54 & 0.41 & 0.36 & 0.31 & 0.24 & 0.16 & 0.09 & 0.06 & 0.03 \\
 & 0.8 & 0.54 & 0.41 & 0.36 & 0.31 & 0.24 & 0.15 & 0.09 & 0.06 & 0.03 \\
 & 1 & 0.55 & 0.41 & 0.36 & 0.31 & 0.24 & 0.15 & 0.09 & 0.06 & 0.03 \\
 & 1.2 & 0.55 & 0.41 & 0.36 & 0.31 & 0.24 & 0.15 & 0.09 & 0.06 & 0.03 \\
 & 1.4 & 0.55 & 0.41 & 0.36 & 0.31 & 0.24 & 0.15 & 0.09 & 0.06 & 0.03 \\
 & 1.6 & 0.55 & 0.41 & 0.36 & 0.31 & 0.24 & 0.15 & 0.09 & 0.06 & 0.03 \\
 & 1.8 & 0.56 & 0.41 & 0.36 & 0.31 & 0.24 & 0.15 & 0.09 & 0.06 & 0.03 \\
 & 2 & 0.56 & 0.41 & 0.36 & 0.31 & 0.24 & 0.16 & 0.09 & 0.06 & 0.03 \\ \hline
\multirow{10}{*}{Secondary} & 0.2 & 0.50 & 0.37 & 0.35 & 0.30 & 0.22 & 0.14 & 0.08 & 0.05 & 0.01 \\
 & 0.4 & 0.52 & 0.38 & 0.35 & 0.30 & 0.22 & 0.14 & 0.08 & 0.05 & 0.01 \\
 & 0.6 & 0.52 & 0.39 & 0.35 & 0.30 & 0.22 & 0.14 & 0.08 & 0.05 & 0.01 \\
 & 0.8 & 0.53 & 0.39 & 0.35 & 0.30 & 0.22 & 0.14 & 0.08 & 0.05 & 0.01 \\
 & 1 & 0.53 & 0.39 & 0.35 & 0.30 & 0.22 & 0.14 & 0.08 & 0.05 & 0.01 \\
 & 1.2 & 0.53 & 0.39 & 0.35 & 0.30 & 0.22 & 0.14 & 0.08 & 0.05 & 0.01 \\
 & 1.4 & 0.53 & 0.39 & 0.35 & 0.30 & 0.22 & 0.14 & 0.08 & 0.05 & 0.01 \\
 & 1.6 & 0.53 & 0.39 & 0.35 & 0.30 & 0.22 & 0.14 & 0.08 & 0.05 & 0.01 \\
 & 1.8 & 0.53 & 0.39 & 0.35 & 0.30 & 0.22 & 0.14 & 0.08 & 0.05 & 0.01 \\
 & 2 & 0.53 & 0.39 & 0.35 & 0.30 & 0.22 & 0.14 & 0.08 & 0.05 & 0.01 \\ \hline
\end{tabular}
\end{sidewaystable}

\begin{sidewaystable}[]
\caption{First principle strain values are reported here for each primary and secondary parameter of density. The strains are reported at various percentiles (maximum, 99th, 95th, 75th, 25th, 5th, 1st, and minimum) as well as the median strain.}
\label{Table: Strain Density}
\begin{tabular}{ccccccccccc}
\hline
\textbf{CDT Classification} & \textbf{Density$^\dagger$} & \textbf{max} & \textbf{99\%} & \textbf{95\%} & \textbf{75\%} & \textbf{median} & \textbf{25\%} & \textbf{5\%} & \textbf{1\%} & \textbf{min} \\ \hline
\multirow{11}{*}{Primary} & 1.0 & 0.71 & 0.46 & 0.36 & 0.28 & 0.23 & 0.14 & 0.08 & 0.05 & 0.02 \\
 & 1.2 & 0.63 & 0.44 & 0.34 & 0.28 & 0.23 & 0.15 & 0.09 & 0.06 & 0.02 \\
 & 1.4 & 0.57 & 0.42 & 0.35 & 0.28 & 0.23 & 0.16 & 0.09 & 0.06 & 0.04 \\
 & 1.6 & 0.57 & 0.42 & 0.35 & 0.28 & 0.23 & 0.16 & 0.09 & 0.06 & 0.04 \\
 & 1.8 & 0.59 & 0.41 & 0.34 & 0.29 & 0.24 & 0.16 & 0.10 & 0.07 & 0.04 \\
 & 2.0 & 0.55 & 0.40 & 0.35 & 0.30 & 0.23 & 0.15 & 0.09 & 0.06 & 0.04 \\
 & 2.2 & 0.49 & 0.39 & 0.34 & 0.30 & 0.24 & 0.16 & 0.10 & 0.07 & 0.04 \\
 & 2.4 & 0.47 & 0.38 & 0.35 & 0.30 & 0.24 & 0.16 & 0.10 & 0.07 & 0.04 \\
 & 2.6 & 0.47 & 0.38 & 0.35 & 0.30 & 0.24 & 0.16 & 0.10 & 0.07 & 0.04 \\
 & 2.8 & 0.46 & 0.38 & 0.35 & 0.30 & 0.24 & 0.16 & 0.10 & 0.07 & 0.04 \\
 & 3.0 & 0.47 & 0.38 & 0.35 & 0.30 & 0.24 & 0.16 & 0.09 & 0.07 & 0.04 \\ \hline
\multirow{10}{*}{Secondary} & 1 & 0.55 & 0.38 & 0.35 & 0.29 & 0.23 & 0.15 & 0.09 & 0.05 & 0.02 \\
 & 2 & 0.55 & 0.38 & 0.34 & 0.30 & 0.23 & 0.15 & 0.09 & 0.06 & 0.02 \\
 & 3 & 0.49 & 0.36 & 0.34 & 0.29 & 0.22 & 0.14 & 0.08 & 0.05 & 0.01 \\
 & 4 & 0.47 & 0.37 & 0.34 & 0.29 & 0.23 & 0.14 & 0.08 & 0.06 & 0.01 \\
 & 5 & 0.47 & 0.37 & 0.35 & 0.29 & 0.22 & 0.14 & 0.08 & 0.05 & 0.01 \\
 & 6 & 0.46 & 0.37 & 0.35 & 0.29 & 0.22 & 0.13 & 0.08 & 0.05 & 0.01 \\
 & 7 & 0.46 & 0.37 & 0.35 & 0.29 & 0.22 & 0.13 & 0.08 & 0.05 & 0.01 \\
 & 8 & 0.44 & 0.37 & 0.35 & 0.29 & 0.22 & 0.13 & 0.08 & 0.05 & 0.02 \\
 & 9 & 0.44 & 0.37 & 0.35 & 0.29 & 0.22 & 0.13 & 0.08 & 0.05 & 0.02 \\
 & 10 & 0.45 & 0.37 & 0.35 & 0.28 & 0.21 & 0.13 & 0.08 & 0.05 & 0.01 \\ \bottomrule
\multicolumn{8}{l}{$^\dagger$Note: Primary density units are ${\rm chords/cm}$, and secondary density units are ${\rm chords/cm}^2$.} 
\end{tabular}
\end{sidewaystable}

\begin{sidewaystable}[]
\caption{First principle strain values are reported here for each primary and secondary parameter of the radius of branching in the CDT. The strains are reported at various percentiles (maximum, 99th, 95th, 75th, 25th, 5th, 1st, and minimum) as well as the median strain.}
\label{Table: Strain Radius}
\begin{tabular}{ccccccccccc}
\hline
\textbf{CDT Classification} & \textbf{Radius (mm)} & \textbf{max} & \textbf{99\%} & \textbf{95\%} & \textbf{75\%} & \textbf{median} & \textbf{25\%} & \textbf{5\%} & \textbf{1\%} & \textbf{min} \\ \hline
\multirow{8}{*}{Primary} & 0.4 & 0.57 & 0.44 & 0.35 & 0.29 & 0.23 & 0.15 & 0.09 & 0.05 & 0.02 \\
 & 0.6 & 0.53 & 0.43 & 0.35 & 0.29 & 0.23 & 0.15 & 0.09 & 0.05 & 0.02 \\
 & 0.8 & 0.52 & 0.41 & 0.35 & 0.30 & 0.23 & 0.15 & 0.09 & 0.05 & 0.03 \\
 & 1.0 & 0.52 & 0.41 & 0.35 & 0.30 & 0.23 & 0.15 & 0.09 & 0.05 & 0.03 \\
 & 1.2 & 0.52 & 0.39 & 0.35 & 0.30 & 0.23 & 0.15 & 0.09 & 0.06 & 0.03 \\
 & 1.4 & 0.51 & 0.39 & 0.35 & 0.30 & 0.23 & 0.15 & 0.09 & 0.06 & 0.03 \\
 & 1.6 & 0.55 & 0.39 & 0.35 & 0.30 & 0.23 & 0.15 & 0.09 & 0.06 & 0.04 \\
 & 1.8 & 0.55 & 0.40 & 0.34 & 0.30 & 0.23 & 0.16 & 0.09 & 0.07 & 0.04 \\ \hline
\multirow{14}{*}{Secondary} & 0.4 & 0.50 & 0.37 & 0.34 & 0.30 & 0.23 & 0.15 & 0.08 & 0.05 & 0.02 \\
 & 0.6 & 0.50 & 0.36 & 0.34 & 0.30 & 0.22 & 0.14 & 0.08 & 0.05 & 0.01 \\
 & 0.8 & 0.49 & 0.36 & 0.34 & 0.30 & 0.22 & 0.14 & 0.08 & 0.05 & 0.01 \\
 & 1.0 & 0.49 & 0.36 & 0.34 & 0.29 & 0.22 & 0.14 & 0.08 & 0.05 & 0.01 \\
 & 1.2 & 0.49 & 0.36 & 0.34 & 0.29 & 0.22 & 0.14 & 0.08 & 0.05 & 0.02 \\
 & 1.4 & 0.48 & 0.36 & 0.34 & 0.29 & 0.22 & 0.14 & 0.08 & 0.05 & 0.02 \\
 & 1.6 & 0.49 & 0.36 & 0.34 & 0.29 & 0.22 & 0.14 & 0.08 & 0.05 & 0.02 \\
 & 1.8 & 0.48 & 0.36 & 0.34 & 0.29 & 0.22 & 0.14 & 0.08 & 0.05 & 0.02 \\
 & 2.0 & 0.48 & 0.36 & 0.34 & 0.29 & 0.21 & 0.14 & 0.08 & 0.05 & 0.02 \\
 & 2.2 & 0.47 & 0.36 & 0.34 & 0.29 & 0.21 & 0.13 & 0.08 & 0.05 & 0.02 \\
 & 2.4 & 0.48 & 0.36 & 0.34 & 0.29 & 0.21 & 0.13 & 0.08 & 0.05 & 0.02 \\
 & 2.6 & 0.47 & 0.36 & 0.33 & 0.29 & 0.21 & 0.14 & 0.09 & 0.06 & 0.03 \\
 & 2.8 & 0.47 & 0.36 & 0.33 & 0.29 & 0.21 & 0.14 & 0.09 & 0.06 & 0.03 \\
 & 3.0 & 0.48 & 0.36 & 0.33 & 0.29 & 0.21 & 0.14 & 0.09 & 0.06 & 0.03 \\ \hline
\end{tabular}
\end{sidewaystable}

\begin{sidewaystable}[]
\caption{First principle strain values are reported here for each primary and secondary parameter of the branching length in the CDT. The strains are reported at various percentiles (maximum, 99th, 95th, 75th, 25th, 5th, 1st, and minimum) as well as the median strain.}
\label{Table: Strain Length}
\begin{tabular}{ccccccccccc}
\hline
\textbf{CDT Classification} & \textbf{Length (mm)} & \textbf{max} & \textbf{99\%} & \textbf{95\%} & \textbf{75\%} & \textbf{median} & \textbf{25\%} & \textbf{5\%} & \textbf{1\%} & \textbf{min} \\ \hline
\multirow{10}{*}{Primary} & 1 & 0.59 & 0.42 & 0.34 & 0.29 & 0.23 & 0.15 & 0.09 & 0.05 & 0.02 \\
 & 2 & 0.54 & 0.40 & 0.35 & 0.30 & 0.23 & 0.15 & 0.09 & 0.06 & 0.03 \\
 & 3 & 0.53 & 0.40 & 0.35 & 0.30 & 0.23 & 0.15 & 0.09 & 0.06 & 0.04 \\
 & 4 & 0.51 & 0.40 & 0.35 & 0.30 & 0.23 & 0.15 & 0.09 & 0.06 & 0.04 \\
 & 5 & 0.51 & 0.40 & 0.35 & 0.30 & 0.23 & 0.15 & 0.09 & 0.06 & 0.04 \\
 & 6 & 0.52 & 0.40 & 0.35 & 0.30 & 0.23 & 0.15 & 0.09 & 0.06 & 0.04 \\
 & 7 & 0.52 & 0.40 & 0.35 & 0.30 & 0.23 & 0.15 & 0.09 & 0.06 & 0.04 \\
 & 8 & 0.52 & 0.40 & 0.35 & 0.30 & 0.23 & 0.15 & 0.09 & 0.06 & 0.03 \\
 & 9 & 0.52 & 0.40 & 0.35 & 0.30 & 0.23 & 0.15 & 0.09 & 0.06 & 0.03 \\
 & 10 & 0.52 & 0.40 & 0.35 & 0.30 & 0.23 & 0.15 & 0.09 & 0.06 & 0.03 \\ \hline
\multirow{10}{*}{Secondary} & 1 & 0.55 & 0.38 & 0.35 & 0.29 & 0.23 & 0.15 & 0.09 & 0.05 & 0.02 \\
 & 2 & 0.55 & 0.38 & 0.34 & 0.30 & 0.23 & 0.15 & 0.09 & 0.06 & 0.02 \\
 & 3 & 0.49 & 0.36 & 0.34 & 0.29 & 0.22 & 0.14 & 0.08 & 0.05 & 0.01 \\
 & 4 & 0.47 & 0.37 & 0.34 & 0.29 & 0.23 & 0.14 & 0.08 & 0.06 & 0.01 \\
 & 5 & 0.47 & 0.37 & 0.35 & 0.29 & 0.22 & 0.14 & 0.08 & 0.05 & 0.01 \\
 & 6 & 0.46 & 0.37 & 0.35 & 0.29 & 0.22 & 0.13 & 0.08 & 0.05 & 0.01 \\
 & 7 & 0.46 & 0.37 & 0.35 & 0.29 & 0.22 & 0.13 & 0.08 & 0.05 & 0.01 \\
 & 8 & 0.44 & 0.37 & 0.35 & 0.29 & 0.22 & 0.13 & 0.08 & 0.05 & 0.02 \\
 & 9 & 0.44 & 0.37 & 0.35 & 0.29 & 0.22 & 0.13 & 0.08 & 0.05 & 0.02 \\
 & 10 & 0.45 & 0.37 & 0.35 & 0.28 & 0.21 & 0.13 & 0.08 & 0.05 & 0.01 \\ \hline
\end{tabular}
\end{sidewaystable}

\newpage

\end{appendices}
\newpage

%% BioMed_Central_Bib_Style_v1.01

\end{document}